\documentclass[aps,pre,twocolumn,groupedaddress,superscriptaddress,nofootinbib,colorlinks,bookmarks=false,citecolor=darkspringgreen,linkcolor=darkolivegreen,urlcolor=navyblue]{revtex4-2}

\usepackage{graphicx}
\usepackage{amsfonts}
\usepackage{amssymb}
\usepackage{amsmath}
\usepackage{dcolumn}
\usepackage[normalem]{ulem}
\usepackage{enumerate}
\usepackage{bbold}
\usepackage{cancel}
\usepackage{mathtools}
\usepackage{siunitx}
\usepackage{amsmath}
\usepackage{mdframed}
\usepackage{dsfont}
\usepackage[caption=false]{subfig}
\usepackage{soul}
\usepackage{orcidlink}
\usepackage{physics}
\usepackage{stackrel}
\usepackage{algorithm}
\usepackage{algpseudocode}

\definecolor{darkgreen}{RGB}{0, 159, 117}
\definecolor{navyblue}{rgb}{0.0, 0.0, 0.5}
\definecolor{darkolivegreen}{rgb}{0.33, 0.42, 0.18}
\definecolor{darktangerine}{rgb}{1.0, 0.66, 0.07}
\definecolor{darkspringgreen}{rgb}{0.09, 0.45, 0.27}
\definecolor{plum}{rgb}{0.45, 0.29, 0.49}

\usepackage{ mathrsfs }

\newcommand{\p}{\hat{p}}
\newcommand{\x}{\hat{x}}

\newcommand{\header}[1]{\noindent {\bf #1.---}}

\newcommand{\papertitle}{Persistent Entanglement Generation in Quantum Machian Fractons}

\begin{document}

\title{\papertitle}

\author{Uddhav Sen}
\affiliation{
Centre for Fluid and Complex Systems, Coventry University, Coventry, CV1 2TT, United Kingdom}

\author{Chris Hooley}
\affiliation{
Centre for Fluid and Complex Systems, Coventry University, Coventry, CV1 2TT, United Kingdom}

\author{Sascha Wald}
\email{sascha.wald@coventry.ac.uk}
\affiliation{
Centre for Fluid and Complex Systems, Coventry University, Coventry, CV1 2TT, United Kingdom}

\begin{abstract}
\noindent 

We show that kinematic mobility constraints, such as those arising in fractonic matter, can persistently generate quantum entanglement even in the absence of conventional interaction potentials.
Using a quantum Machian fracton model initialized in separable coherent states, we demonstrate that the constraint entangles particles even while observables initially still match classical expectations.
Subsequently, observable expectation values strongly deviate from their classical behavior.
This is accompanied by an entanglement burst and sharp density peaks forming at the constraint edge, as the finite extent of the wave function triggers arrest well before the classical trajectories slow down.
Even after the burst subsides and the observables arrest, unbounded squeezing enforced by the kinematic constraint drives continued entanglement growth.
Our work identifies a mechanism rooted in squeezing and non-Gaussianity from configuration-dependent mobility constraints that drives entanglement beyond dynamical arrest and connects directly to Hilbert space fragmentation and slow thermalization in many-body dipole-conserving systems.
\end{abstract}

\maketitle

\begin{figure}[t]
    \centering
    \includegraphics[width=.9\columnwidth]{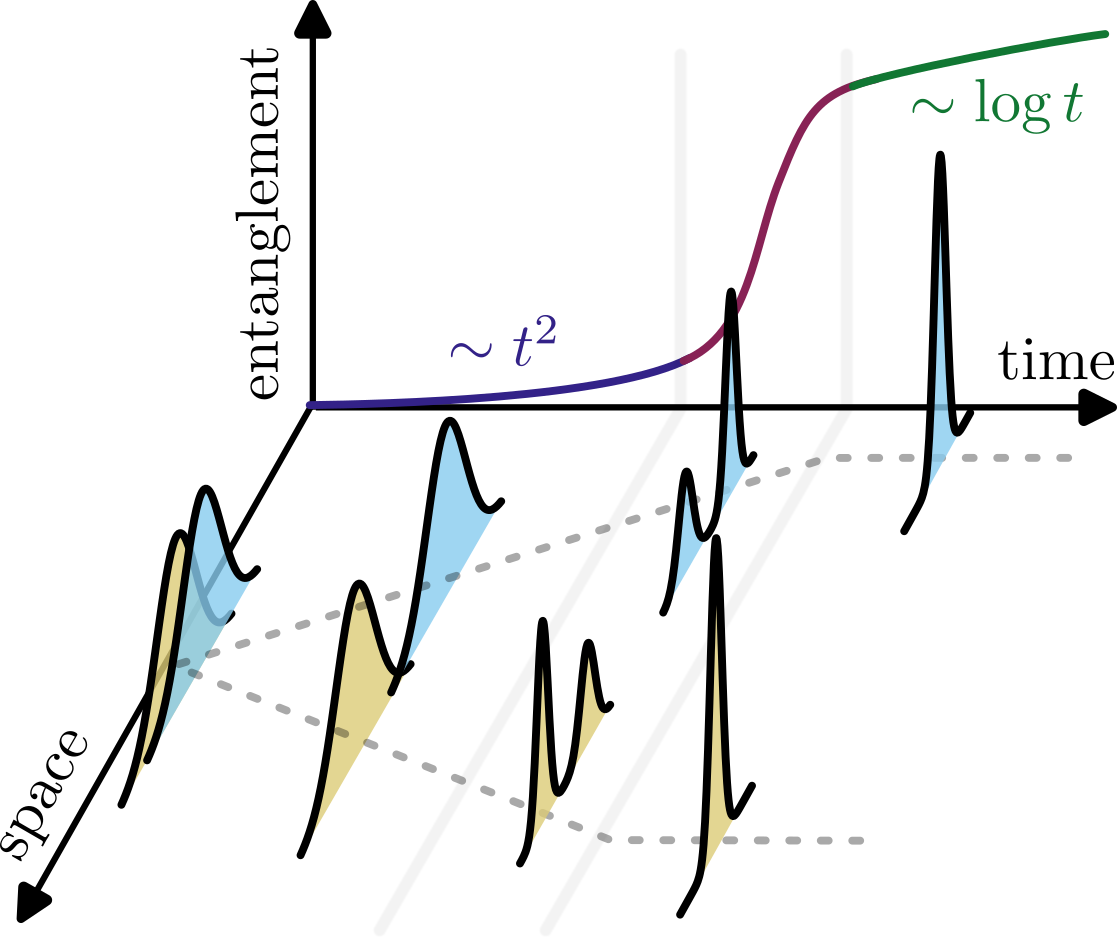}
    \caption{{\bf Dynamics and entanglement of quantum Machian fractons.}  The main plot sketches the evolution of the single-particle probability densities of two quantum Machian fractons (blue and yellow), starting from an initial product state of two Gaussian wave packets.  
    The back plane shows a sketch of the entanglement entropy between the two fractons as a function of time with three distinct regimes.
    Initially, the entanglement grows slowly (purple), followed by a rapid growth during the `two peak' phase of the evolution (red); however, it continues to grow even when the peaks of the wave packets are no longer moving (green).}
    \label{fig:sketch}
\end{figure}

\header{Introduction}Fractonic matter is characterized by restricted mobility, where individual particles are immobilized by subsystem or multipole symmetries while their bound composites may move freely~\cite{Chamon2005, Haah2011, Vijay2016, Pretko2017, Nandkishore2019, Gromov2024}.  
Excitations with restricted mobility, later named fractons~\cite{Vijay2016}, first appeared in exactly solvable three-dimensional lattice models, introduced to study quantum glassiness~\cite{Chamon2005} and self-correcting quantum memory~\cite{Haah2011, Yoshida2013}.
In these models, the fractonic behavior stems from subsystem
symmetries, while in the continuum it follows from the exact conservation of
charge together with its multipole moments~\cite{Pretko2017,Pretko2018}. 
Classifications of fractonic
phases have been proposed in terms of the algebraic structure of these
moments~\cite{Pretko2017, Pretko2018, Gromov2019, Seiberg2020}.
Here we show that this type of restricted mobility leaves a distinctive imprint on
entanglement, which continues to grow long after the particles have ceased to
move (Fig.~\ref{fig:sketch}).

Fractonic physics is by now an experimentally accessible frontier, with practical implementations actively engineered across diverse physical platforms.
Kinetic constraints and the resulting non-ergodic fracton dynamics have been observed in ultracold atoms using tilted optical lattices~\cite{Scherg2021, Kohlert2019, Guardado2020}, while programmable Rydberg atom arrays offer synthetic quantum platforms capable of preparing non-Abelian fracton orders~\cite{Verresen2021, Bluvstein2022}.
Furthermore, fractonic dynamics can be physically emulated in custom-designed, macroscopic `fractolectric' circuits that naturally extend to the quantum regime~\cite{Pretko2019}.  

Despite this experimental progress, the development of a universal continuum or hydrodynamic description of fractonic motion remains a central theoretical challenge~\cite{Gromov2020, Lucas2020, Glorioso2023, prakash2024a}. 
A `Machian' scheme, in which a particle's inertia is determined by its position
relative to the other constituents of the system, was proposed in
ref.~\cite{prakash2024a} and has since been developed
further~\cite{prakash2024b, Babbar2025, Howes2025, Sadki2026a, singh2026,sadki2026c}.
This approach introduces a position-dependent mass that diverges when particles separate beyond a threshold distance, enforcing dipole conservation and preserving locality of the model.

While the classical mechanics of such Machian systems has recently been established, quantizing such motion requires careful consideration of the position-dependent mass~\cite{BenDaniel1966, Delgado2019, Costa2021}.  
Recently, the spectral properties of continuous dipole-conserving quantum fractons were explored in the few-body regime~\cite{Sadki2026b}.
However, the effects of the restricted mobility on the entanglement between the particles, which may have important applications in quantum information processing, remains unexplored.

In this Letter, we analyze the mutual entanglement of quantum Machian fractons.
For an initially unentangled particle pair, we show that the growth of entanglement exhibits three successive regimes (Fig.~\ref{fig:sketch}):\ a `free fractonic' regime, for which we calculate the entanglement entropy analytically; a `burst' regime, characterized by rapid entanglement growth as the wavefunction weight is transferred to the point where the mass diverges; and a `frozen' regime, during which the entanglement entropy continues to grow even though the expectation values of the particles' positions have plateaued. 

\header{Model}We consider two particles moving in one dimension with positions $x_j$ and momenta $p_j$ ($j=1,2$), subject to a dynamics that conserves the dipole moment $D = x_1 + x_2$ and the total momentum $P = p_1 + p_2$. 
(The general form of the dipole moment is $D = Q_1 x_1 + Q_2 x_2$; in this work, we set the particles' charges $Q_1 = Q_2 = 1$.)

Ref.~\cite{prakash2024a} proposed a local classical Hamiltonian that generates such dipole-conserving dynamics:
\begin{align}\label{eq:H_cl}
 \mathscr{H} = \frac{1}{2} B_a(x_1 - x_2) (p_1 - p_2)^2,
\end{align}
with $B_a(x) =\Theta(a-|x|)$ their chosen form of the spatially dependent inverse mass.  Note that this inverse mass is zero for $\vert x \vert > a$, where $a$ is a positive constant.
 
Here, we consider a quantum extension of this classical model. 
Specifically, we study two distinguishable quantum particles in one spatial dimension, the observables for which are their respective position and momentum operators $\{\x_j,\p_j\}$, $j=1,2$. 
We take these to obey canonical commutation relations, $[ \x_j,\p_k ] = i \delta_{jk}$ and $[ \x_j,\x_k ] = [ \p_j,\p_k ] = 0$. (Here and henceforth we set $\hbar \equiv 1$.)
We extend the classical Hamiltonian (\ref{eq:H_cl}) to a Hermitian quantum Hamiltonian: 
\begin{align}
\hat{H} = \frac{\p_1 - \p_2}{\sqrt{2}} B_a(\x_1 - \x_2)\frac{\p_1 - \p_2}{\sqrt{2}}.
\label{eq:H_q}
\end{align}

We may write this more compactly by changing variables to the center-of-mass frame, viz.\ $\hat{q}_\pm = (\x_1 \pm \x_2)/\sqrt{2}$ and $\hat{\pi}_\pm = (\p_1 \pm \p_2)/\sqrt{2}$.  
These new coordinates and momenta obey the canonical commutation relations $[\hat{q}_{\mu},\hat{\pi}_\nu] = i \delta_{\mu\nu}$, $[ \hat{q}_\mu,\hat{q}_\nu ] = [ \hat{\pi}_\mu,\hat{\pi}_\nu ] = 0$, where $\mu,\nu = \pm$; in terms of them, the Hamiltonian (\ref{eq:H_q}) becomes
\begin{align}
    \hat{H} = \hat{\pi}_- B_a\big(\sqrt{2}\,\hat{q}_-\big) \hat{\pi}_-.
\end{align}

Since ${\hat H}$ depends only on the relative momentum ${\hat \pi}_{-}$ and the relative coordinate ${\hat q}_{-}$, it appears that we can consider only the relative motion of the particles, thereby reducing the problem to an effective one-particle system.
However, this requires that the initial quantum state of the two particles be separable in this coordinate system, which is not guaranteed.
Here we ensure it by considering both particles to be independently initialized in a coherent state at the origin, so that the two-particle wavefunction $\Psi(x_1,x_2;t)$ at time $t=0$ is
\begin{equation}
    \Psi(x_1,x_2;0) = \psi_{\sigma,k}(x_1) \psi_{\sigma,-k}(x_2),
\end{equation}
where
\begin{align}
    \psi_{\sigma,k}(x_j) = 
    \left(\frac{\gamma}{\pi }\right)^{1/4}
    \exp\left( - \frac{x_j^2}{2\sigma^2} + i k x_j \right),
    \label{eq:gauss}
\end{align}
with $\gamma = \Re(\sigma^{-2})$.
The choice of equal and opposite momenta for the two particles ensures that the dipole moment is conserved.
In the ultracold-atom context, for example, this particle configuration can be readily achieved using harmonic traps to prepare the quantum states~\cite{Morinaga1999,Kaufman2012,Thompson2013,Kaufman2014,Xin2021,Brown2023}.

Crucially, the initial state we chose also factorizes in center-of-mass coordinates:
\begin{align}
\Psi(x_1,x_2;0) dx_1 dx_2 &=  \psi_{\sigma',0}(q_+)\psi_{\sigma',k'}(q_-)dq_+dq_-, 
\label{eq:state_trafo}
\end{align}
with $\sigma' = \sigma/2$ and $k'=\sqrt{2}\,k$.
Since the Hamiltonian depends neither on the center-of-mass coordinate $q_+$ nor on the center-of-mass momentum $\pi_+$, that part of the wavefunction does not evolve in time.  
We may thus focus solely on the dynamics of the relative coordinate $q_-$.
This problem can be mapped to a standard particle-in-a-box problem, albeit with Neumann rather than Dirichlet boundary conditions (see Supplemental Material for further details~\cite{SM}).
This allows us to
obtain the eigenenergies $E_n = 2 n^2  \pi^2/a^2$, $n\in \tfrac{1}{2}\mathbb{N}$ and the corresponding eigenfunctions 
\begin{equation}
\varphi_n(q_-)=
\left\{ \begin{array}{lll}
\displaystyle \sqrt{\frac{2}{a}} \sin\left(\frac{\sqrt{2} \, n\pi q_-}{a}\right),& \qquad & n=\frac{1}{2},\frac{3}{2},\frac{5}{2},\ldots;\\
& \\
\displaystyle \sqrt{\frac{2}{a}} \cos\left(\frac{\sqrt{2}\,n\pi q_-}{a}\right),& & n=1,2,3,\ldots.
\end{array} \right.
\end{equation}

\begin{figure}[t]
    \centering
    \includegraphics[width=\columnwidth]{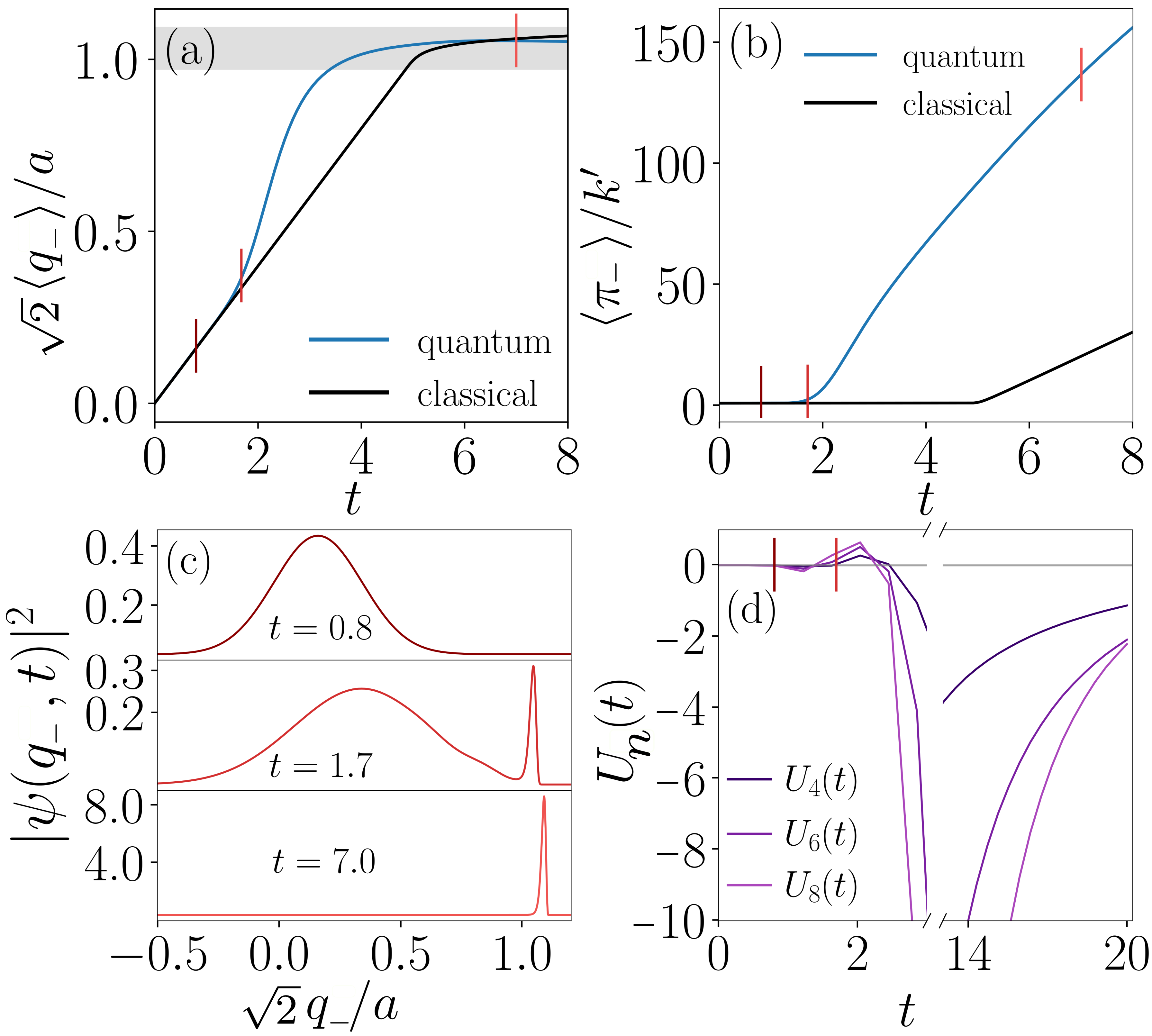}
    \caption{{\bf Dynamics of quantum fractons.} (a) The time-evolution of the expectation value of the relative co-ordinate $\langle q_-\rangle$ (blue) compared to the classical case (black). (b) The same for the relative momentum, $\langle \pi_{-}\rangle$. (c) The time-evolution of the probability density of the relative coordinate $q_-$.  (d) Binder-cumulant-like parameters of the probability density, $U_n$, as a function of time.  For a Gaussian distribution all $U_n = 0$. The three vertical lines in Figs. (a), (b) and (d) in varying shades of crimson indicate different snapshots in time shown in Fig. (c).
    }
    \label{fig:fig2}
\end{figure}

\header{Dynamical properties}We begin with a numerical analysis of the time-evolution of the relative wavefunction $\psi(q_-)$ and of the observables $\langle {\hat q}_{-} \rangle$ and $\langle {\hat \pi}_{-} \rangle$.
To avoid numerical difficulties due to the discontinuous nature of the inverse mass $B_a(x)$, we replace it by a softened one, $ B_{a,\eta}(x)= (\tanh\eta(x+a)-\tanh\eta(x-a))/2$, such that $\lim_{\eta\to\infty} B_{a,\eta}(x) = B_a(x)$. 
For the numerical results presented in the main text we use $\eta = 10$, $a=5$ and the initial state parameters $\sigma' = k' = 1$.
In Fig.~\ref{fig:fig2}(a) and (b) we present numerical results for $\langle \hat{q}_-\rangle$ and $\langle\hat{\pi}_-\rangle$ in this quantum fractonic problem, in comparison with the corresponding classical quantities.
The quantum results are obtained by applying a coordinate and field rescaling to the time-dependent Schr{\"o}dinger equation to bring it into a form with a position-independent mass, and then carrying out the time-evolution using a standard finite-difference method, see Supplemental Material~\cite{SM}.

As we see in Fig.~\ref{fig:fig2}(a), the distance between the particles initially increases with time in perfect agreement with its classical counterpart.
However, well before the point at which the classical distance reaches its limiting value, the classical and quantum curves separate from each other.
This discrepancy arises from the finite spread of the quantum wavepacket: the tail of $\psi(q_-)$ probes the varying separation-dependent mass before the peak of the packet enters the region of significant variation.
This effect is amplified by spreading of the wavepacket as it propagates.  At long times both the classical and quantum expectation values of the interparticle distance tend to a value in the interval $[a/\sqrt{2} - \eta^{-1}, a/\sqrt{2}+\eta^{-1}]$, which corresponds to the broadened step in the softened box function.
The relative momentum shows a similar behavior, see Fig.~\ref{fig:fig2}(b).  Initially the classical and quantum values remain constant, at times $t \gtrsim 2$ the quantum value starts to increase much faster than the classical one, which again is a wavefunction-tail effect (see below).  In the long-time limit, in both the classical and quantum cases, the expectation value of the relative momentum of the particles increases without limit.

To explore the origin of the separation of the classical and quantum curves at intermediate times we consider the probability density derived from the relative wave function, shown in Fig.~\ref{fig:fig2}(c). At early times (upper subpanel, deep crimson) the Gaussian shape of the coherent state is preserved.  At intermediate times (middle subpanel, crimson red) a sharp second peak appears at the boundary $\sqrt{2} q_- \approx a$, even though the original peak has not yet reached that position.  At late times (lower subpanel, light red) the probability density has been almost entirely transferred to this boundary peak.

We may quantify the bimodal nature of the probability density at intermediate times using the Binder-cumulant-like parameters $U_n$, defined by
\begin{equation}
U_n = 1- \frac{\langle (q_- - \langle q_-\rangle)^n \rangle}{(n-1)!!\langle (q_- - \langle q_-\rangle)^2 \rangle^{n/2}}
\end{equation}
where $n$ is a positive even integer. 
These have the property that, for a Gaussian distribution, $U_n = 0$ for all $n > 2$.  These parameters are plotted in Fig.~\ref{fig:fig2}(d) for $n=4,6,8$.

We see that indeed they remain near zero at early times, when the fractonic evolution is free; this agrees with an analytical treatment in the $a \to \infty$ limit, which shows that $\psi(x,t) \simeq \psi_{\sigma(t),k'}(x-k' t)$, up to a global phase, with $\sigma(t) = \sqrt{\sigma'^2 + it}$ (see Supplemental Material \cite{SM} for details).

However, at intermediate times, the higher-order parameters $U_4$, $U_6$, and $U_8$ all become very large and negative.  
This corresponds to the development of the sharp boundary peak in the probability density, a phenomenon that has no classical counterpart. 
At late times, the wave packet becomes strongly localized at the edge due to the fractonic constraint; in this regime, the parameters $U_4$, $U_6$, and $U_8$ again approach zero, implying that the boundary peak, once formed, becomes more Gaussian at late times.

\begin{figure}[t]
    \centering
    \includegraphics[width=\columnwidth]{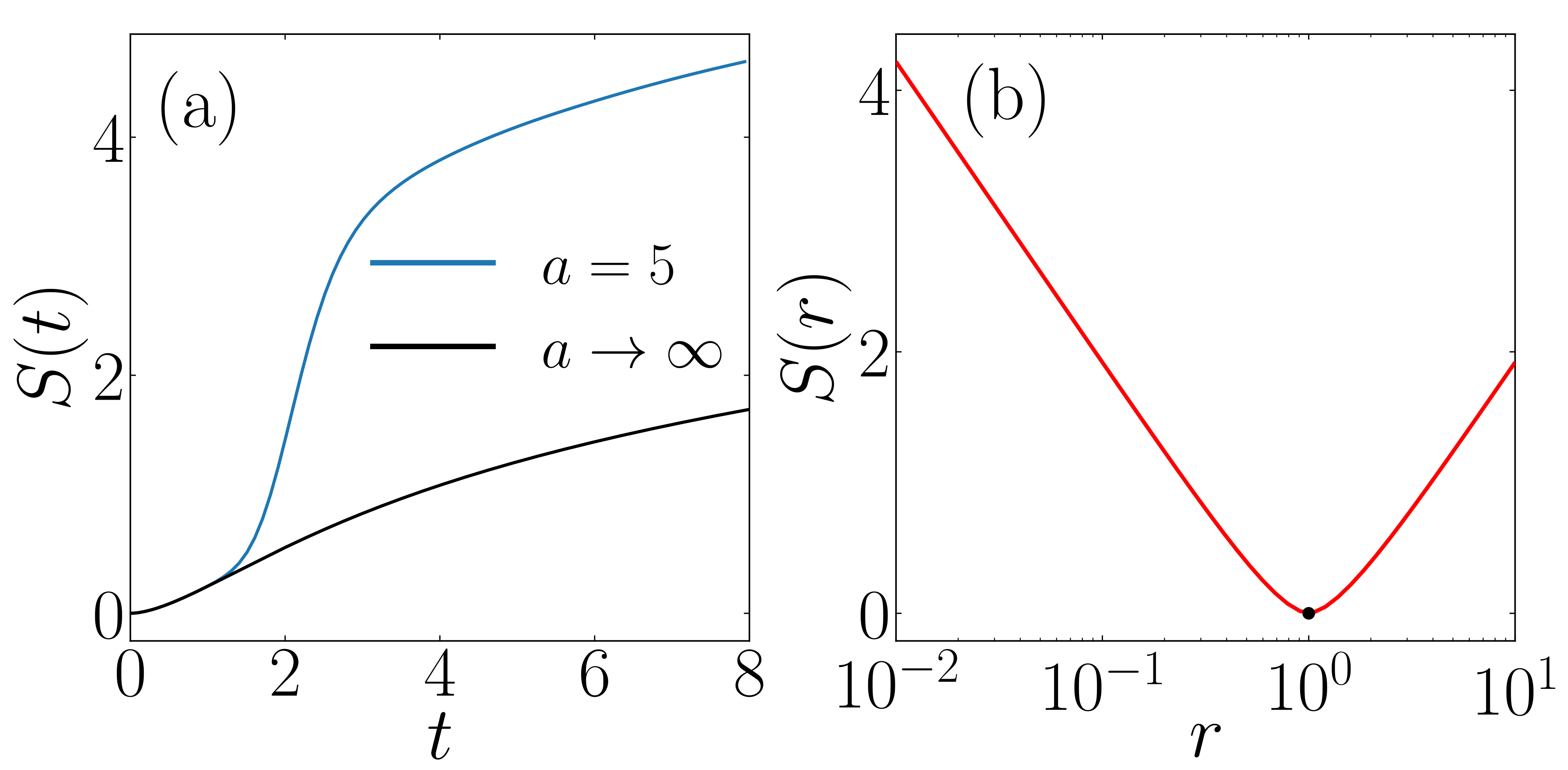}
    \caption{{\bf Entanglement entropy growth of quantum Machian fractons.}
    (a) The single-particle entanglement entropy $S(t)$ for a finite box function width $a$ (blue curve) compared to the `free' fractonic evolution obtained in the $a \to \infty$ limit (black curve).
    (b) The single-particle entanglement entropy for time-independent but non-separable states in which the relative-coordinate Gaussian is squeezed relative to the center-of-mass-coordinate one. The black marker highlights the zero-entropy point at $r = 1$, corresponding to a separable state.
}
    \label{fig:ent}
\end{figure}

\header{Entanglement}A key feature distinguishing quantum Machian fractons from their classical counterparts is the generation of entanglement.
The corresponding entanglement entropy $S$ is obtained from the one-particle reduced density matrix
\begin{equation}
    \rho_1(x_1,x_1') = \int dx_2\, \Psi(x_1,x_2)\Psi^*(x_1',x_2).
\end{equation}
Here, the degrees of freedom of the second particle are traced out and the entanglement entropy is given by $S= -\mathrm{Tr}[\rho_1\log\rho_1]$.
Figure~\ref{fig:ent}(a) shows the time evolution of $S$, revealing persistent entanglement generation throughout the fractonic quantum dynamics.
Even during the initial `free' fractonic evolution, prior to the development of the boundary peak, the particles entangle due to their correlated motion.
We see that the initial growth is followed by a burst regime during which the particles become strongly entangled in a short time window.
Subsequently, the entanglement generation slows down, but crucially never stops.
We shall now provide analytic arguments for the form of the entanglement entropy growth in the short-time (`free' fractonic evolution) and long-time regimes.

First, we note that whenever the relative wave function $\psi(q_-)$ is Gaussian, the entanglement entropy is analytically accessible.
Indeed, we show in the Supplemental Material~\cite{SM}  that, in such cases, the entanglement entropy is obtained from the symplectic eigenvalue $\nu$ of the covariance matrix of the single-mode Gaussian state:
\begin{align}
    S = \biggl(\nu + \frac{1}{2} \biggr)\log(\nu + \frac{1}{2}) - \left(\nu - \frac{1}{2}\right)\log(\nu - \frac{1}{2}). \label{seq}
\end{align}
As we have seen in the previous section, the relative wave function is well approximated by a Gaussian both during the initial `free' fractonic regime and at late times.

In the `free' fractonic regime the position-dependence of the inverse mass is negligible, and  expectation values of observables agree with classical predictions.
This regime can be studied in the limit $a\to\infty$, where the variance of the relative wavefunction evolves as $\sigma_-^2(t) = \sigma'^2 + i t$.
This yields the symplectic eigenvalue~\cite{Williamson36, Holevo99, Simon2000, Weedbrook2012}
\begin{equation}
\nu(t) = \frac{1}{2}\sqrt{1 + \frac{t^2}{4 (\sigma')^4}}.
\end{equation}
Inserting this into (\ref{seq}) and Taylor-expanding around $t=0$ shows that $S(t) \sim t^2$ at early times.
We see that this analytical prediction is in perfect agreement with the early-time numerical evaluation.
However, the `free' evolution fails to capture the entanglement burst, which implies that the burst is intimately related to the squeezing of the wavefunction in the region where the inverse mass varies rapidly.

At late times the growth of the entanglement entropy appears to be qualitatively captured again by the analytical approach. 
This can be understood from our findings in the previous section, where we have shown that the boundary peak becomes more Gaussian at late times, see Fig.~\ref{fig:fig2}(d).
We therefore approximate the relative wave function as a Gaussian $\psi_{\sigma_{-},0}(q_- - a/\sqrt{2})$  with real width $\sigma_-$.
This leads to the symplectic eigenvalue (see Supplemental Material~\cite{SM})
\begin{equation}
\nu(r) = \frac{\sigma_+^2 + \sigma_{-}^2}{4 \sigma_+ \sigma_-} = \frac{1}{4} \left( r + \frac{1}{r} \right),
\label{eq:sm_nu_static}
\end{equation}
with $\sigma_+$ being the width of the center-of-mass wave function and $r= \sigma_-/\sigma_+$.
However, as we see from Fig.~\ref{fig:fig2}(c), the relative-coordinate Gaussian is progressively squeezed, i.e.\ $\sigma_{-}\to 0$.
In this limit, the entanglement entropy behaves as $S\sim -\log r$.
Although we do not have access to the exact functional form of the late-time variance $\sigma_-$, we can rule out an exponential behavior on the grounds of Fig.~\ref{fig:fig2}(d), since an exponential would lead to a much quicker localization on the edge.
Similarly, a logarithmic behavior can be ruled out since there is no time-scale competition, which leaves only a polynomial behavior, i.e.\ $r\sim t^{-\alpha}$.
This yields an entanglement entropy that continues to grow logarithmically, $S \sim \log t$, at late times.

\header{Conclusion}We have taken a first step towards a quantum information
theory of continuum Machian fractons, showing for a pair of particles that
restricted mobility alone, without any interaction potential, acts as a
persistent source of entanglement.
In the free fractonic regime the dipole constraint holds the center of mass
fixed while the relative coordinate spreads, so that the relative momentum grows
faster than in the classical model, and an initially unentangled pair entangles
immediately, with an entropy that grows quadratically at short times and
logarithmically at late times.
Between these regimes, the entanglement generation is significantly accelerated as the wavefunction weight is transferred to a second peak at the edge of the kernel, which constitutes a strongly non-Gaussian process with no classical counterpart.
Once the separation has frozen, energy conservation forces the relative momentum
to diverge as the mobility kernel is suppressed, so the relative wavepacket is
squeezed without limit and the entanglement entropy continues to grow
logarithmically even though the position expectation values have plateaued.

Entanglement generation thus outlives the dynamical arrest, driven by a mechanism generic to configuration-dependent, dipole-conserving continuum dynamics rather than specific to the functional form of our mobility kernel. 
This mechanism extends beyond the two-body scenario:
the wave function undergoes strong squeezing and develops pronounced non-Gaussian components in the process.
Therefore, our study directly relates to Hilbert space fragmentation and to the slow thermalization of dipole-conserving systems.
From a technological point of view, though, it may well be that the `burst' region that we observe at intermediate times will be the most significant.  There are already well known examples of quantum effects leading to rapid (superlinear) changes in conventional observables, e.g.\ the phenomenon of superradiance \cite{GROSS1982301}, which has found application in quantum battery protocols \cite{RevModPhys.96.031001}.  This model shows a similar superlinear increase in entanglement, which might find application in quantum state preparation.  Exploration of this possibility would be a worthy topic of future research.

\header{Data Availability}The data displayed in the figures is available on Zenodo~\cite{zenodo}.

\header{\bf Acknowledgment}This work made use of the ARCHER2 UK National Supercomputing Service (https://www.archer2.ac.uk) as well as Coventry University's EPYC high performance computer.

\bibliography{refs}
\bibliographystyle{apsrev4-1}


\setcounter{equation}{0}
\setcounter{figure}{0}
\setcounter{table}{0}
\makeatletter
\renewcommand{\theequation}{S\arabic{equation}}
\renewcommand{\thefigure}{S\arabic{figure}}
\makeatletter
\onecolumngrid
\newpage

\setcounter{page}{1}
\begin{center}
{\Large SUPPLEMENTAL MATERIAL}
\end{center}
\begin{center}
\vspace{0.8cm}
{\Large \papertitle }
\end{center}
 \begin{center}
 Uddhav Sen,$^{1}$
 Chris Hooley,$^{1}$ and
 Sascha Wald$^{1}$
 \end{center}
 \begin{center}
 $^1${\em Centre for Fluid and Complex Systems, Coventry University, Coventry, CV1 2TT, United Kingdom}
 \end{center}

\section*{Boundary Matching Conditions}
Here we derive the boundary matching condition of the wave function of the relative degree of freedom $q_-$ at $q_-=\pm \frac{a}{\sqrt{2}}$ where the inverse mass function changes abruptly.
We therefore explicitly integrate the time-independent Schr\"odinger equation across $q_- = \pm\frac{a}{\sqrt{2}}$.
We recall the Schr\"odinger equation reads
\begin{align}
i \partial_t \psi(q_-,t) = -\partial_{q_-} \left( B_a(\sqrt{2}q_-) \partial_{q_-} \psi(q_-,t) \right).
\end{align}
First, we separate variables with the Ansatz $\psi(q_-,t) = \varphi(q_-) f(t)$ which yields the eigenvalue equation
\begin{align}
-\frac{d}{dq_-} \left( B_a(\sqrt{2}q_-) \frac{d}{dq_-} \varphi({q_-}) \right) = E \varphi({q_-}).
\label{eq:stat_ev}
\end{align}
To establish the matching condition across ${q_-} = \frac{a}{\sqrt{2}}$, we integrate~(\ref{eq:stat_ev}) over an infinitesimal interval $\left[\frac{a}{\sqrt{2}}-\epsilon, \frac{a}{\sqrt{2}}+\epsilon\right]$ of width $2\epsilon$ and subsequently take the limit $\epsilon \to 0$.
\begin{align}\label{alpha1}
- \left( B_a(\sqrt{2}(a/\sqrt{2}+\epsilon))  \varphi'(a/\sqrt{2}+\epsilon) - B_a(\sqrt{2}(a/\sqrt{2}-\epsilon))  \varphi'(a/\sqrt{2}-\epsilon) \right) = E \int_{\frac{a}{\sqrt{2}}-\epsilon}^{\frac{a}{\sqrt{2}}+\epsilon} \varphi(q_-) dq_-
\end{align}
Assuming that $\varphi$ is continuous across the boundary, the right-hand side vanishes in the limit $\epsilon\to0$, whereas the left-hand side evaluates to $\varphi'\left(\frac{a}{\sqrt{2}}^-\right) = \varphi'\left(\frac{a}{\sqrt{2}}^+\right)/M$, with $M$ being a regularizer in the infinite-mass regime.
In the limit $M \to \infty$, the right-hand side vanishes for any finite gradient $\varphi'\left(\frac{a}{\sqrt{2}}^+\right)$, enforcing the derivative condition $\varphi'\left(\frac{a}{\sqrt{2}}^-\right) = \varphi'\left(-\frac{a}{\sqrt{2}}^+\right) = 0$.

\section*{Sturm--Liouville Operator Representation}
In this section we will show that the dynamics of the relative degree of freedom $q_-$ can be mapped to a standard problem of a particle in a potential.
We note that the time-independent Schr\"odinger equation~(\ref{eq:stat_ev}) is of the Sturm-Liouville form and use in the remainder of this section the softened inverse mass function $ B_{a,\eta}$ since our goal is to obtain a numerically tractable differential equation.
First, we introduce the transformation $y(q_-) = \int_{0}^{q_-} dz /\sqrt{ B_{a,\eta}(z)}$ such that
\begin{align}
   dy = \frac{dq_-}{\sqrt{ B_{a,\eta}(q_-)}} 
\end{align}
which allows us to carry out the coordinate transform of the full differential equation.
Next, we introduce the rescaled wave function $\phi(y) = [ B_{a,\eta}(q_-)]^{1/4} \psi(q_-)$ such that $\int dy |\phi(y)|^2 = 1$ follows from the normalization of $\psi$.
The transformed time-independent Schr\"odinger equation thus reads
\begin{equation}
-\frac{1}{2}\frac{d^2\phi(y)}{dy^2} +  V_{\text{eff}}(y)  \phi(y) = E \phi(y)
\end{equation}
with the effective potential contribution
\begin{align}
    V_{\text{eff}}[q_-(y)] = \frac{1}{8}  B_{a,\eta}''(q_-(y)) - \frac{1}{32} \frac{[ B_{a,\eta}'(q_-(y))]^2}{ B_{a,\eta}(q_-(y))}.
\end{align}
The described transform also trivially applies to the time-dependent Schr\"odinger equation which yields the dynamical equation
\begin{align}
-i\partial_t \phi(y,t) = -\frac{1}{2}\phi''(y,t)+V_{\text{eff}}(y)\phi(y,t).
\end{align}

\section*{Relative wave function in the free fractonic limit}
In this section, we consider the dynamics of the relative degree of freedom $q_-$ in the limit $a\rightarrow\infty$.
In this limit, the relative motion is free with the corresponding Hamiltonian $\hat{H} = (\hat{\pi}_-)^2 /(2m)$.
We consider an initial Gaussian state localized at the origin with relative momentum $k'$ and width $\sigma'$, viz. $\Psi(q_-,0) = \psi_{\sigma',k'}$, cf. Eq.~(\ref{eq:gauss}).
To determine the time evolution, we Fourier transform the Schr\"odinger equation of the relative degree of freedom such that the time evolution reduces to accumulating a dynamical phase factor.
We find
\begin{equation}
\Psi(q_-, t) = \frac{\sigma'^{1/2}}{\pi^{1/4} (\sigma^2(t))^{1/2}} \exp\left[ -\frac{(q_-)^2}{2 \sigma^2(t)} + \frac{i k' {\sigma'^2}}{\sigma^2(t)}\left(q_- - \frac{k' t}{2m}\right) \right],
\label{eq:supp_psi_exact}
\end{equation}
with the time dependent variance $\sigma^2(t)=\sigma'^2+it/m$.
This allows us to verify the relative position and momentum expectation values $\langle q_-(t) \rangle = k' t/m$, and $\langle \pi_-(t) \rangle = k'$.
These expectation values are consistent with two particles moving apart from each other with constant momentum.
We may also evaluate the second-order moments $\langle (q_-)^2 \rangle = \sigma'^2 \left(1/2 + t^2/(2m^2\sigma'^4)\right)+{k'^2} t^2/m^2$  and $\langle  (\pi_-)^2 \rangle =  k'^2 + 1/(2\sigma'^2)$.
This yields
\begin{align}\label{uncer}
    \Delta q_- \Delta \pi_- &=  \frac{1}{2} \sqrt{1 + \frac{t^2}{m^2\sigma'^4}}
\end{align}

\section*{Single-Particle Entanglement Entropy in the Anisotropic Free Limit}
In this section, we will characterize the entanglement between the two quantum particles, subject to the dynamical constraint, in the case in which the dynamics does not freeze, i.e. $a\to \infty$.
Since the initial state is separable in the center-of-mass and relative coordinate, and the dynamics only evolves the relative-coordinate wave function, the state of the system remains separable in this coordinate system at all times.
Further, to make this problem tractable, we consider the relative wavefunction to be Gaussian which is a valid description in the early- and the late-time regime.
The center-of-mass wave function remains Gaussian throughout.
Thus, the states we consider here can be written as $\Psi(q_+, q_-) = \psi_{\sigma_+, 0}(q_+) \, \psi_{\sigma_-, k'}(q_-)$, cf. Eq.~(\ref{eq:gauss}).
Although $\Psi(q_+,q_-)$ is separable at all times, $\Psi(x_1,x_2)$ is not, cf.~(\ref{eq:state_trafo}).
We proceed to quantify the entanglement between the particles by considering the von Neumann entropy of the single particle reduced density matrix, viz.
\begin{equation}
\rho(x_1, x_1') = \int_{-\infty}^{\infty} dx_2 \, \Psi(x_1, x_2) \, \Psi^*(x_1', x_2).
\label{eq:sm_rho1_def}
\end{equation}
Since the two-particle wave function is Gaussian, the reduced density matrix also remains Gaussian.
Any single-mode Gaussian state is completely characterized by its first moments and its covariance matrix $V$, viz.
\begin{align}
V&=
\begin{pmatrix}
\langle(\Delta x_1)^2\rangle & \operatorname{Re}\langle\Delta x_1\,\Delta p_1\rangle\\
\operatorname{Re}\langle\Delta x_1\,\Delta p_1\rangle & \langle(\Delta p_1)^2\rangle
\end{pmatrix},
\label{eq:sm_cov_def}
\end{align}
with $\Delta O = O-\langle O\rangle$.
First moments represent local phase-space displacement operations which can be eliminated via unitary Weyl translations $D = \exp(i( \langle p_1 \rangle x_1 -  \langle x_1 \rangle p_1))$. 
Since local unitary operations preserve the spectrum of the reduced density matrix, first moments do not contribute to quantum entanglement.
Any real, positive-definite symmetric covariance matrix $V$ can be diagonalized via a symplectic matrix $M$~\cite{Williamson36}.
Hence we write
\begin{equation}
V = M^T \begin{pmatrix} \nu & 0 \\ 0 & \nu \end{pmatrix} M,
\end{equation}
where $\nu \ge 1/2$ is the symplectic eigenvalue of $V$. 
The value of $\nu$ is a canonical invariant under phase-space symplectic transformations and is determined directly by $\nu = \sqrt{\det V}$.
The Gaussian single-particle reduced density matrix $\rho$ is unitarily equivalent to an effective thermal state of a single-mode quantum harmonic oscillator, $\rho_{\mathrm{th}} = Z^{-1} \exp(-\beta \hat{a}^\dagger \hat{a})$, with mean thermal occupation number $\bar{n} = \nu - 1/2$~\cite{Williamson36,Banchi15}.
Since one-particle unitary transformations do not affect the two-particle entanglement, the entropy of the reduced single-particle state is identical to the one of the corresponding thermal state $\rho_{\mathrm{th}}$. 
We obtain the entropy for a single-mode Gaussian state
\begin{equation}
S(\nu) = \left( \nu + \frac{1}{2} \right) \log \left( \nu + \frac{1}{2} \right) - \left( \nu - \frac{1}{2} \right) \log \left( \nu - \frac{1}{2} \right).
\label{eq:sm_gaussian_entropy}
\end{equation}
We can now evaluate the entanglement entropy for $\Psi(x_1,x_2)$ in terms of the centre-of-mass and relative width parameters $\sigma_{\pm}$.
Generally, $\sigma_+$ is real, while $\sigma_-$ is initially real but becomes complex during the time evolution. 
It is therefore convenient to introduce $\gamma = \operatorname{Re}(1/\sigma_-^2)$, and $\delta = \operatorname{Im}(1/\sigma_-^2)$,
in terms of which we can explicitly evaluate the covariance matrix, viz.,
\begin{equation}
V = \frac{1}{4\gamma}
\begin{pmatrix} 
1 + \gamma\sigma_+^2 & -\delta \\
-\delta & \gamma/\sigma_+^2 + \gamma^2 + \delta^2 
\end{pmatrix}.
\end{equation}
From here, we straightforwardly find the symplectic eigenvalue
\begin{equation}
\nu = \sqrt{\det V} = \frac{|\sigma_+^2 + \sigma_-^2|}{4 \sigma_+ \sqrt{\operatorname{Re}(\sigma_-^2)}}.
\label{eq:sm_nu_master}
\end{equation}

\end{document}